\documentclass[twocolumn,prc,amssymb, aps,superscriptaddress, 
amsmath]{revtex4}

\usepackage{color}
\usepackage{amsmath,mathtools,booktabs,
            microtype} 

\usepackage{graphicx}
\usepackage{dcolumn}
\usepackage{bm}
\usepackage{multirow}
\usepackage{times}
\usepackage{url}
\usepackage{tikz}
\usepackage[version=4]{mhchem}

\usetikzlibrary{arrows,decorations.pathmorphing}
\usetikzlibrary{arrows.meta}
\usetikzlibrary{calc}

\definecolor{violet}{HTML}{602969}
\definecolor{red}{HTML}{FC0009}
\definecolor{orange}{HTML}{FF6319}
\definecolor{green}{HTML}{00933C}
\definecolor{blue}{HTML}{0036A6}
\definecolor{yellow}{HTML}{FFBE00}
\definecolor{lightgrey}{HTML}{A7A9AC}

\newcommand{\be}{\begin{equation}}
\newcommand{\ee}{  \end{equation}}
\newcommand{\ba}{\begin{eqnarray}}
\newcommand{\ea}{  \end{eqnarray}}

\begin{document}

\title{Thermalization of closed chaotic many-body quantum systems}

\author{Hans A. \surname{Weidenm\"uller}}
\email{haw@mpi-hd.mpg.de}
\affiliation{Max-Planck-Institut f\"ur Kernphysik, Saupfercheckweg 1, D-69117 Heidelberg, Germany}

\date{\today}

\begin{abstract}A closed quantum system thermalizes if for time $t \to
  \infty$, the function ${\rm Tr} (A \rho(t))$ tends asymptotically to
  ${\rm Tr} (A \rho_{\rm eq})$. Here $A$ is an operator that
  represents an observable, $\rho(t)$ is the time-dependent density
  matrix, and $\rho_{\rm eq}$ its equilibrium value. We investigate
  thermalization of a chaotic many-body quantum system by combining
  the Hartree-Fock (HF) approach and the Bohigas-Giannoni-Schmit (BGS)
  conjecture. The HF Hamiltonian defines an integrable system and the
  gross fatures of the spectrum. The residual interaction locally
  mixes the HF eigenstates. The BGS conjecture implies that the
  statistics of the resulting eigenvalues and eigenfunctions agrees
  with random-matrix predictions. In that way, the Hamiltonian $H$ of
  the system acquires statistical features. The agreement of the
  statistics with random-matrix properties is local, i.e, confined to
  an interval $\Delta$ (the correlation width). With $\rho(t) = \exp
  \{ - i t H / \hbar \} \rho(0) \exp \{ i H t / \hbar \}$, the
  statistical properties of $H$ define the statistical properties of
  ${\rm Tr} (A \rho(t))$.

  We define an ensemble of Hamiltonians. The ensemble comprises the
  original Hamiltonian $H$ as an element. All members of the ensemble
  have the same statistical properties as $H$. We calculate mean value
  and correlation function of ${\rm Tr}(A \rho(t))$ as averages over
  the ensemble. In the semiclassical regime, the average $\langle {\rm
    Tr}(A \rho(t)) \rangle$ decays with time on the scale $\hbar /
  \Delta$ towards an asymptotic value. If the energy spread of the
  system is of order $\Delta$, that value is given by ${\rm Tr} (A
  \rho_{\rm eq})$, and the statistical fluctuations of ${\rm Tr}(A
  \rho(t))$ (measured in terms of the correlation function) around the
  mean value vanish. That shows that for every member of the ensemble
  (including the original Hamiltonian), ${\rm Tr}(A \rho(t))$
  thermalizes.

  The correlation width $\Delta$ is the central parameter of our
  approach. It defines the interval within which the spectral
  fluctuations agree with random-matrix predictions. It defines the
  maximum energy spread of the system that permits thermalization. And
  it defines the time scale $\hbar / \Delta$ within which ${\rm Tr}(A
  \rho(t))$ approaches the asymptotic value ${\rm Tr}(A \rho_{\rm
    eq})$. We argue that the correlation width $\Delta$ occurs
  generically in chaotic quantum systems where it plays the same
  central role.
  
\end{abstract}

\maketitle

\section{Introduction}
\label{int}

Thermalization addresses the time evolution of the expectation value
${\rm Tr} (A \rho(t))$ of a classical observable (represented in
Hilbert space by a Hermitean operator $A$) in a closed chaotic quantum
system ${\cal S}$ with density matrix $\rho(t)$. Thermalization
postulates that asymptotically (time $t \to \infty$) the expectation
value ${\rm Tr} (A \rho(t))$ of $A$ tends toward its equilibrium
value,
\ba
\label{i1}
{\rm Tr} (A \rho(t)) \to {\rm Tr} (A \rho_{\rm eq}) \ .
\ea
Here $\rho_{\rm eq}$ is the density matrix of ${\cal S}$ in
statistical equilibrium. The asymptotic relation~(\ref{i1}) is
expected to hold in the semiclassical regime of very high excitation
energy $E$ and for systems with very small energy spread $\delta E \ll
E$. Thermalization is seen as the analogue in quantum statistical
mechanics of ergodicity in classical statistical
mechanics. Reviews are given in Refs.~\cite{DAl16, Aba19}.

Starting with Refs.~\cite{Neu29, Sre99}, thermalization is commonly
derived~\cite{DAl16, Aba19} in the eigenbasis of the Hamiltonian $H$
of ${\cal S}$ with states labeled $\alpha, \beta$ where
\ba
\label{i2}
{\rm Tr} (A \rho(t)) = \sum_{\alpha \beta} A_{\alpha \beta}
\rho_{\beta \alpha}(t) \ .
\ea
It is assumed that the matrix elements of the operator $A_{\alpha
  \beta}$ are random variables while the matrix elements $\rho_{\beta
  \alpha}(t)$ of the density matrix are taken to be fixed (i.e.,
non-statistical). If the statistical properties of $A_{\alpha \beta}$
reflect the statistical properties of the Hamiltonian $H$, that
assumption is not convincing. Consistency requires that both
$A_{\alpha \beta}$ and $\rho_{\beta \alpha}(t)$ be taken as stochastic
variables. That was demonstrated in Ref.~\cite{Wei23}. The
distribution of $A_{\alpha \beta}$ and $\rho_{\beta \alpha}(t)$ should
follow from the spectral fluctuation properties of the Hamiltonin $H$
of the system ${\cal S}$ (i.e., from the joint probability
distribution of eigenvalues and eigenfunctions). The approach to
thermalization presented here is built upon that premise. We focus
attention on a chaotic many-body quantum system ${\cal S}$. Using a
combination of a mean-field approach and the Bohigas-Giannoni-Schmit
conjecture~\cite{Boh84}, we determine the generic spectral fluctuation
properties of the Hamiltonian of ${\cal S}$. That makes it possible to
calculate ${\rm Tr} (A \rho(t))$ explicitly. We thereby determine the
time scale of thermalization. We quantify the conditions ``high
excitation energy'' and $\delta E \ll E$ that are required for
thermalization. We show what happens to the asymptotic
relation~(\ref{i1}) if the system is not in the semiclassical regime,
or if the condition $\delta E \ll E$ is violated.

The present paper follows Ref.~\cite{Wei23} wherein a random-matrix
approach to thermalization has been developed. We closely follow the
overall layout of that paper. Except for occasional references we do
not, however, repeat here all of the central arguments nor some of the
technical steps of Ref.~\cite{Wei23}.

The paper is structured as follows. In Section~\ref{app} we describe
our approach. In Section~\ref{hart} we formulate the Hartree-Fock
approach and determine the generic statistical properties of the
Hamiltonian of a chaotic many-body quantum system ${\cal S}$. These
are used to define in Section~\ref{stat} a random Hamiltonian
ensemble. In terms of that ensemble, ${\rm Tr} (A \rho(t))$ turns into
a time-dependent stochastic process. In Sections~\ref{ens} and
\ref{cor} we calculate ensemble average and correlation function,
respectively, of ${\rm Tr} (A \rho(t))$. In Section~\ref{vio} we
extend the treatment to time-reversal noninvariant
systems. Section~\ref{dis} contains a summary and the discussion.

\section{Approach}
\label{app}

With $H$ the Hamiltonian, the density matrix is given by
\ba
\label{a1}
\rho(t) = \exp \{ - i H t / \hbar \} \Pi \exp \{ i H t / \hbar \} \ .
\ea
Here $\Pi$ is the statistical operator of ${\cal S}$. The operator
$\Pi$ describes the probability distribution of ${\cal S}$ over the
states in Hilbert space. Normalization implies
\ba
\label{a2}
{\rm Tr} (\Pi) = 1 \ .
\ea
All eigenvalues $\pi_\kappa$, $\kappa = 1, 2, \ldots$ of $\Pi$ are
positive or zero. Together with Eq.~(\ref{a2}), that implies $0 \leq
\pi_\kappa \leq 1$ and $\sum_\kappa \pi_\kappa = 1$. The system ${\cal
  S}$ being closed, the operator $\Pi$ is independent of time. The
entire time dependence of ${\rm Tr} (A \rho(t))$ is due to the unitary
time-evolution operator $\exp \{ - i H t / \hbar \}$ and its Hermitean
adjoint in Eq.~(\ref{a1}). Thermalization is expected to occur
irrespective of the initial conditions on ${\cal S}$, i.e., for any
choice of the statistical operator $\Pi$ provided only that the system
is in the semiclassical regime and $\delta E \ll E$ applies. In
Ref.~\cite{Sre99} it is assumed, for instance, that $\Pi$ possesses
only a single nonzero eigenvalue and, thus, describes a system very
far from statistical equilibrium. The operator $A$ may be constrained
by the condition that $A$ possesses a viable classical counterpart.

As stressed in the literature~\cite{DAl16, Aba19}, thermalization is
not universal. Integrable quantum systems and systems that show
many-body localization do not thermalize~\cite{DAl16}. Thermalization
is expected to occur generically, however, in chaotic many-body
systems~\cite{DAl16}. From Eq.~(\ref{a1}) it is not evident how that
difference in behavior comes about. Indeed, writing the observable
${\rm Tr} (A \rho(t))$ explicitly in the eigenbasis of $H$ with
orthonormal eigenvectors denoted by the letters $\alpha, \beta,
\ldots$ and eigenvalues labeled $E_\alpha, E_\beta, \ldots$ and using
the Einstein summation convention we have
\ba
\label{a3}
{\rm Tr} (A \rho(t)) = A_{\alpha \beta} \exp \{ - i E_\beta t / \hbar \}
\Pi_{\beta \alpha} \exp \{ i E_\alpha t / \hbar \} \ .      
\ea
That expression is universal and holds irrespective of the particular
dynamical properties of the system. The matrix elements $A_{\alpha
  \beta}$ and $\Pi_{\beta \alpha}$ and the diagonal elements
$\rho_{\alpha \alpha}$ of the density matrix in Eq.~(\ref{a3}) are
independent of time, while the nondiagonal elements $\rho_{\beta
  \alpha}(t)$ with $\beta \neq \alpha$ carry the phase factors $\exp
\{ i (E_\alpha - E_\beta) t / \hbar \}$ and, thus, keep oscillating in
time forever. These statements, too, hold alike for integrable,
chaotic, and for systems that show many-body localization.

The fundamental difference between chaotic systems on the one hand and
integrable systems or systems that show many-body localization on the
other, lies in the spectral fluctuation properties of $H$. For chaotic
quantum systems the Bohigas-Giannoni-Schmit (BGS) conjecture states
that these coincide locally with those of the random-matrix ensemble
in the same symmetry class (orthogonal, unitary, or
symplectic)~\cite{Boh84}. No such correspondence exists for integrable
systems or for systems that show many-body localization, and
statistical assumptions cannot be used there. Therefore, we expect
Eq.~(\ref{i1}) to follow from the statistical properties of the
Hamiltonian of a chaotic quantum system. That is in line with the
standard approach to thermalization~\cite{Sre99, DAl16, Aba19}. As
mentioned in Section~\ref{int}, in that approach it is shown that
thermalization follows and can be worked out explicitly if it is
assumed that the elements $A_{\alpha \beta}$ of the observable $A$ in
Eq.~(\ref{a2}) are Gaussian random variables while the matrix elements
$\Pi_{\beta \alpha}$ of the statstical operator are taken to be
nonstatistical. For reasons given in Section~\ref{int} and, in more
detail, in Ref.~\cite{Wei23} we follow here a different route.

Using an arbitrary fixed basis of states labeled $(\mu, \nu, \ldots)$
in Hilbert space we write Eq.~(\ref{a3}) in the form
\ba
\label{a4}
{\rm Tr} (A \rho(t)) = A_{\mu \nu} U_{\nu \rho}(t) \Pi_{\rho \sigma}
U^*_{\mu \sigma}(t) \ .
\ea
Here $U(t)$ is the unitary time-evolution operator
\ba
\label{a5}
U(t) = \exp \{ - i H t / \hbar \} \ .
\ea
To work out thermalization we consider the matrices $A_{\mu \nu}$ and
$\Pi_{\rho \sigma}$ as fixed, i.e., independent of time and
nonstatistical. The entire time dependence and the statistical
properties of ${\rm Tr} (A \rho(t))$ reside in $U(t)$ and its
Hermitean adjoint. We show that the statistical properties of the
Hamiltonian of the chaotic many-body quantum system ${\cal S}$ and,
thus, of $U(t)$ can be determined via a combination of a mean-field
approach and of the BGS conjecture~\cite{Boh84}. As mentioned above,
the conjecture states that the spectral fluctuation properties of the
Hamiltonian $H$ of a chaotic quantum system coincide locally with
those of the random-matrix ensemble in the same symmetry class. The
word ``locally'' refers to the existence of an energy interval
$\Delta$. The interval $\Delta$ limits the range of energies wherein
the specral fluctuation properties of a chaotic quantum system
coincide with those of random-matrix theory. The interval $\Delta$
occurs generically in chaotic quantum systems. It is called the
Thouless energy in diffusive quantum systems~\cite{Alt86}, the
localization length in the theory of banded random
matrices~\cite{Wil91}, it shows up in nuclear shell-model
spectra~\cite{Zel96}, it has been identified for the quantum Sinai
billard~\cite{Arv81} and for a quantized Cantori system~\cite{Win88},
and it is given by the pion mass in chiral random-matrix theory of
quantum chromodynamics~\cite{Ver00}. The existence of $\Delta$ is an
integral part of the BGS conjecture. Without it, the Hamiltonian would
be a member of the Gaussian Orthogonal Ensemble (GOE) of random
matrices. That is physically impossible since that ensemble carries no
information. In what follows we refer to $\Delta$ as the width of
random-matrix correlations, in short the correlation width. In nuclear
physics the expression ``spreading width'' is often used
instead~\cite{Zel96}. We do not follow that custom because it refers
to wave-function mixing only, not to eigenvalue statistics.

The BGS conjecture forms a cornerstone of our approach. That is
consistent with the central role played by the conjecture in the
theory of chaotic quantum systems, see, for instance, the book by
Haake~\cite{Haa10}. The conjecture has been tested successfully on a
vast variety of numerical models of classically chaotic systems, and
in the low-energy regime of several physical systems. We cite
Refs.~\cite{Win86, Del86, Sie91} as examples. While an analytical
proof of the conjecture is still lacking, its universal validity for
chaotic quantum systems has, for the case of the level-level
correlation function, been demonstrated in a physically convincing
manner with the help of Gutzwiller's periodic orbit theory, see
Refs.~\cite{Mue07} and \cite{Mue11}. Periodic-orbit theory is
semiclassical by construction. That is particularly fitting in the
present context wherein the semiclassical regime plays an important
role. We believe that the BGS conjecture is now universally accepted,
with well-understood exceptions lending additional support to its
generic validity, see, for instance, Ref.~\cite{Elk21}.

\section{Hartree-Fock Approach}
\label{hart}

We establish the generic spectral fluctuation properties of the
Hamiltonian $H$ of a chaotic many-body quantum system by combining the
mean-field approach with the BGS conjecture. In the main body of the
paper, we confine ourselves to time-reversal-invariant
systems. Violation of time-reversal invariance is addressed in
Section~\ref{vio}. Without further mention we focus attention on a
fixed set of conserved quantum numbers. We assume that the many-body
Hamiltonian is governed by two-body interactions. The Hartree-Fock
(HF) approach yields an approximation $H_{\rm HF}$ to $H$. The total
Hamiltonian $H$ can then be written as
\ba
\label{s1}
H = H_{\rm HF} + V
\ea
where $V$ is the residual interaction. The eigenstates of $H_{\rm HF}$
are labeled $m, n, \ldots$, and we have
\ba
\label{s2}
(H_{\rm HF})_{m n} = {\cal E}_m \delta_{m n} \ . 
\ea
The eigenvalues ${\cal E}_m$ of $H_{\rm HF}$ with $m = 1, 2, \ldots$
are ordered so that ${\cal E}_m \leq {\cal E}_n$ for $m < n$. Partial
degeneracies are removed by diagonalization of $V$ in the subspace
spanned by the degenerate HF states. Without changing labels, we then
have ${\cal E}_m < {\cal E}_n$ for all $m < n$. The remaining part of
the residual interaction is again denoted by $V$. Since $H_{\rm HF}$
is integrable, the eigenvalues ${\cal E}_m$ follow a Poisson
distribution. The eigenstates of the HF Hamiltonian are products of
single-particle states or, because of antisymmetrization and/or
angular-momentum coupling, are linear combinations of such product
states. They do not carry any statistical properties. The average
level density $\rho(E)$ is defined as an average over an energy
interval centered on $E$ that contains a large number of eigenvalues
${\cal E}_m$. In a typical many-body system, $\rho(E)$ increases
nearly exponentially with excitation energy $E$.

The residual interaction $V$ is dominated by one- and two-body
forces. That is what we focus on. The one-body matrix elements of $V$
connect many-body HF states that differ only in the occupation of a
single single-particle HF state. The two-body matrix of $V$ connect
only many-body HF states that differ in the occupation of not more
than two single-particle HF states. The one-body matrix elements of
$V$ involve the overlap of two single-particle HF eigenfunctions. The
bigger the difference in quantum numbers characterizing these two
functions, the smaller the overlap. Therefore, there exists an
effective cutoff that limits the number of many-body HF states
connected to a given one by the one-body matrix elements of $V$. The
same arguments apply to the two-body matrix elements of $V$. With the
cutoff, the number of nonvanishing matrix elements of $V$ in each row
and column of the matrix $H$ in Eq.~(\ref{s1}) is finite. The matrix
representing $H$ is sparse and banded. The band width is defined by
the matrix elements of $V$ farthest from the main diagonal. The band
width increases with increasing excitation energy because the average
level density does. The average number of zeros separating neighboring
nonvanishing elements $V$ increases. In the sense of Ref.~\cite{Bra12}
the matrix $H$ is local and is, therefore, a viable candidate for
thermalization.

In spite of its being sparse in the representation of Eq.~(\ref{s1}),
the residual interaction $V$ is capable of mixing the HF states
sufficiently strongly to produce eigenvalues and eigenfunctions that
locally follow random-matrix predictions. That has been demonstrated,
for instance, by extensive numerical calculations of spectra and
eigenfunctions for nuclei in the middle of the $s-d$ shell in
Ref.~\cite{Zel96}. The shell-model Hamiltonian used in
Ref.~\cite{Zel96} is the sum of single-particle energies and a
two-body residual interaction and, therefore, belongs to the class of
Hamiltonians considered in Eq.~(\ref{s1}). Confining Hilbert space to
the states of the $s-d$ shell, the authors diagonalized Hamiltonian
matrices of dimensions up to several thousand, defined by parity,
total spin and isospin. In the centres of the spectra, the results for
eigenfunctions and eigenvalues showed close agreement with GOE
predictions.

At the same time, the results of Ref.~\cite{Zel96} indicate limits of
that agreement. For the eigenvalues, the nearest-neighbor spacing
distribution agrees with the GOE prediction. However, the spectral
rigidity agrees with the logarithmic dependence predicted by the GOE
for small energy intervals only. At some point the spectral rigidity
shows a sudden upbend and increasing disagreement with the GOE result
beyond that point (see, for instance, Fig.~27 of
Ref.~\cite{Zel96}). Likewise, the eigenfunctions of the shell-model
Hamiltonian, although characterized by a Gaussian distribution of
their components, are not a total mixture of all basis
states. Correspondingly, the residual interaction does not spread the
unpertubed many-particle states $m$ of the shell model uniformly over
the eigenfunctions of the shell-model Hamiltonian. Rather, their
distribution is centered at the unperturbed energy ${\cal E}_m$ and
has a finite width $\Delta$ in energy (the ``spreading width''),
smaller than the range of the total spectrum of the shell-model
Hamiltonian. In Ref.~\cite{Zel96}, two forms of the distribution of
the unperturbed states play a role. If the matrix elements of $V$,
although sparse, cover uniformly the HF matrix~(\ref{s1}), the matrix
$V$ is not banded, and the distribution has the shape of a Lorentzian
with width $\Delta$ given by the standard expression
\ba
\label{s3}
\Delta = 2 \pi \langle V^2_{m n} \rangle \rho(E) \ .
\ea
Here $\langle V^2_{m n} \rangle$ denotes the mean square matrix
element, the average with respect to $m$ and $n$ being taken over an
interval centered at $E$ that comprises a sufficiently large number of
HF states labeled $(m, n)$. Actually, Eq.~(\ref{s3}) is a variant of
Fermi's golden rule. Without averaging over $m$ it describes the
spreading of the HF state $| m \rangle$ over the eigenstates of the
system that is due to mixing with the states $n \neq m$. The average
over $m$ is used to attain independence of $\Delta$ of the particular
state $| m \rangle$. Expressions of the form of Eq.~(\ref{s4}) have
been extensively derived, discussed, and used in the literature, see,
for instance, Refs.~\cite{Boh69, DeP11}. If, on the other hand, the
matrix $V$ is banded, the shape of the distribution is Gaussian. The
results of Ref.~\cite{Zel96} are in betwen these two limiting cases,
see Fig.~46. If shells higher than the $s-d$ shell were to be included
in the calculation, the band structure of $V$ would be pronounced more
clearly, and we would expect the distribution to approach the Gaussian
form more closely. 

In summary, the work of Ref.~\cite{Zel96} has shown that the residual
interaction $V$ is capable of producing strong mixing of the basis HF
states that lead to spectral fluctuations of the GOE type. These
fluctuations are local. Their range in energy is bounded. That is true
both for the statistics of eigenvalues and for the statistics of
eigenfunctions. In the present paper we assume that the energy
interval $\Delta$ confining statistical agreement with GOE predictions
is the same for eigenvalues and eigenfunctions. To the best of or
knowledge, that assertion has not been tested yet in realistic cases.

We accordingly use the following formulation of the BGS conjecture. At
each energy $E$, there exists an interval $\Delta$ (called the
correlation width and semiquantitatively given by Eq.~(\ref{s3}))
centered on $E$ within which the action of $V$ in Eq.~(\ref{s1})
causes spectral fluctuations of the GOE type. Such fluctuations exist
also at an energy $E'$ far removed from $E$ (so that $|E - E'| \gg
\Delta$) but are not correlated with the ones around $E$.

We are going to use our assumption in the semiclassical regime. It is,
therefore, important to understand the dependence of $\Delta$ on
excitation energy $E$. A simple scaling argument shows that $\Delta$
as given by Eq.~(\ref{s3}) depends only weakly on $E$. For an energy
$E'$ much bigger than $E$ we account for the exponantial growth of the
level density by the scaling parameter $\alpha \gg 1$ so that $\rho(E)
\to \rho(E') = \alpha \rho(E)$. That causes the number of zeros
separating two adjacent nonvanishing matrix elements of $V$ to grow,
too. The growth is proportional to $\alpha$. In consequence, the
average $\langle V^2_{m n} \rangle$ is multiplied by the factor $1 /
\alpha$, and $\Delta$ remains the same. Although the estimate for the
spreading width of a banded matrix differs from Eq.~(\ref{s3}) it
seems reasonable to expect a qualitatively similar behavior in that
case. We conclude that the correlation width $\Delta$ does not possess
the nearly exponential dependence on energy that characterizes
$\rho(E)$ (although $\Delta$ may display a polynomial dependence on
energy). Therefore, the number of states $N_\Delta = \Delta \rho(E)$
in the interval $\Delta$ increases strongly with energy. In the
semiclassical regime we have $N_\Delta \gg 1$, and we may use an
expansion in inverse powers of $N_\Delta$, keeping only leading-order
terms.

\section{Statistical Properties}
\label{stat}

The Hamiltonian $H$ in Eq.~(\ref{s1}) is diagonalized by an orthogonal
transformation with elements $O_{m \alpha}$ and has eigenvalues
$E_\alpha$, so that
\ba
\label{s4}
H_{m n} = \sum_\alpha O_{m \alpha} E_\alpha O_{n \alpha} \ . 
\ea
As done for the eigenvalues of the HF Hamiltonian, we order the
nondegenerate eigenvalues $E_\alpha$ so that $E_\alpha < E_\beta$ for
$\alpha < \beta$. By assumption, the Hamiltonian $H$ describes a
chaotic many-body quantum system. Therefore, the BGS conjecture holds
in the form formulated in Section~\ref{hart}. Obviously that requires
$V$ to be sufficiently strong so as to cause strong local mixing of
the HF states $| m \rangle$.

According to the BGS conjecture, there exists for each energy $E$ an
energy interval $\Delta$ centered on $E$ within which the eigenvalues
$E_\alpha$ and eigenvectors $O_{m \alpha}$ approximately obey GOE
statistics. The approximation improves with the number $N_\Delta =
\Delta \rho(E)$ of states located within the interval $\Delta$ and
becomes excellent in the semiclassical regime. For convenience we
assume $\Delta {\rm d} \rho(E) / {\rm d}E \ll \rho(E)$ so that in the
interval $\Delta$ the average level density $\rho(E)$ is effectively
constant.

It is technically difficult to use the statistical properties of $H$
directly to work out mean value and variance of ${\rm Tr} (A
\rho(t))$. It is preferable to construct an ensemble of Hamiltonians
that all have the same statistical properties as $H$ and to show that
the resulting expression for the ensemble average of ${\rm Tr} (A
\rho(t))$ holds for every member of the ensemble and, therefore, also
for $H$. That is the route we follow. We construct the ensemble by
focusing attention on HF states with energies ${\cal E}_m$ and
eigenstates of $H$ with eigenvalues $E_\alpha$ that are located within
a correlation length $\Delta$ of each other. In the limit $N_\Delta
\to \infty$, the eigenvalues $E_\alpha$ and matrix elements $O_{m
  \alpha}$ are uncorrelated random variables. The eigenvalues obey
Wigner-Dyson statistics. The matrix elements $O_{m \alpha}$ are
zero-centered real Gaussian random variables with second moments
\ba
\label{s5}
\big\langle O_{m \alpha} O_{m' \alpha'} \big\rangle = \delta_{m m'}
\delta_{\alpha \alpha'} F[({\cal E}_m - \overline{E}_\alpha)^2 / \Delta^2]
\ . 
\ea
The big angular brackets denote the ensemble average. The function
$F(x)$ in Eq.~(\ref{s5}) limits the range of GOE correlations of the
matrix elements $O_{m \alpha}$ in terms of the correlation width
$\Delta$, and $\overline{E}_\alpha$ is the ensemble average of the
$E_\alpha$ with fixed $\alpha$~\footnote{In order to respect the
statistical independence of the random variables $O_{m \alpha}$ and
$E_\alpha$, it is necessary in Eq.~(\ref{s5}) to use
$\overline{E}_\alpha$ rather than $E_\alpha$ itself. Stiffness of the
GOE spectrum implies that the fluctuations of the $E_\alpha$ are of
the order of the mean level spacing $1 / \rho(E)$, and that the mean
value exists. The relation $E_\alpha < E_\beta$ for $\alpha < \beta$
implies $\overline{E}_\alpha < \overline{E}_\beta$.}. Thus, $F(x)$
must be positive for all $x \geq 0$ and must fall off steeply with
increasing $x$ for $x > 1$. Eigenvalues $E_\alpha, E_\beta$ obeying
$|E_\alpha - E_\beta| \gg \Delta$ are taken to be uncorrelated. The
same is true for matrix elements $O_{m \alpha}, O_{n \beta}$ referring
to states with energies ${\cal E}_m, {\cal E}_n$ and $E_\alpha,
E_\beta$ that obey $|{\cal E}_m - {\cal E}_n| \gg \Delta$ or
$|E_\alpha - E_\beta| \gg \Delta$. Eq.~(\ref{s5}) implies that the
residual interaction spreads the HF eigenstates over an energy
interval of width $\Delta$. Conversely, each eigenstate at enery
$E_\alpha$ is a linear combination of HF states located within a
neighborhood of width $\Delta$ centered at $E_\alpha$. We note that
for the GOE itself, the correlations extend over the entire range of
the spectrum. In Eq.~(\ref{s5}) the function $F(x)$ is, thus, replaced
by $1 / N$ where $N$ is the GOE matrix dimension, signalling complete
mixing of all states of the spectrum. With $\lambda$ defining the
width of the GOE spectrum, the role of $F(x)$ is taken over by the GOE
average level density $\sqrt{! - E^2 / (4 \lambda^2)}$. In the
present case, such complete mixing is restricted to states within the
correlation width $\Delta$. While our random-matrix ensemble may not
reflect physical reality very closely near the ground state (where the
inequality $N_\Delta \gg 1$ may not apply), it certainly does apply in
the semiclassical regime with its very high average level density.

The orthogonality relations $\sum_\alpha \langle O_{m \alpha} O_{m'
  \alpha} \rangle = \delta_{m m'}$ and $\sum_m \langle O_{m \alpha}
O_{m \alpha'} \rangle = \delta_{\alpha \alpha'}$ impose restrictions
on $F(x)$. For $N_\Delta \gg 1$ we use the continuum limit, replacing
summations by integrations. Then the orthogonality relations read
\ba
\label{s6}
1 &=& \int {\rm d} \overline{E}_\alpha \ \rho(\overline{E}_\alpha)
F[({\cal E}_m - \overline{E}_\alpha)^2 / \Delta^2] \nonumber \\
&=& \int {\rm d} {\cal E}_m \ \rho({\cal E}_m) F[({\cal E}_m -
  \overline{E}_\alpha)^2 / \Delta^2] \ .
\ea
The range of integration comprises the entire spectrum but actually is
limited by the sharp falloff of $F(x)$ with increasing $x$. We assume
that the strong local mixing of HF states due to $V$ does not affect
the large-scale average level density. Then the average density
$\rho(\overline{E}_\alpha)$ of the average eigenvalues
$\overline{E}_\alpha$ of $H$ equals the average level density
$\rho({\cal E}_m)$ of the HF states ${\cal E}_m$. With $\Delta {\rm d}
\rho(E) / {\rm d}E \ll \rho(E)$ that gives
\ba
\label{s6a}
1 &=& \rho(E) \int {\rm d} \overline{E}_\alpha \
  F[({\cal E}_m - \overline{E}_\alpha)^2 / \Delta^2] \nonumber \\
  &=& \rho(E) \int {\rm d} {\cal E}_m \ F[({\cal E}_m -
  \overline{E}_\alpha)^2 / \Delta^2] \ .
\ea
Here $E$ is an energy somewhere within the interval $\Delta$.

The analytical form of $F(x)$ has ben addressed in
Section~\ref{hart}. For the standard model (which yields Eq.~(\ref{s3})
for the correlation width $\Delta$) the function $F(x)$ has Lorentzian
shape,
\ba
\label{s8}
&& F[({\cal E}_m - \overline{E}_\alpha)^2 / \Delta^2] \\
&& = \frac{1}{2 \pi \rho(E)} \ \frac{\Delta} {({\cal E}_m -
  \overline{E}_\alpha)^2 + (1/4) \Delta^2} \ . \nonumber
\ea
More realistic in the present case is the Gaussian form
\ba
\label{s7}
&& F[({\cal E}_m - \overline{E}_\alpha)^2 / \Delta^2] = \frac{1}
{\sqrt{2 \pi} \rho(E) \Delta} \nonumber \\
&& \qquad \times \exp \{ - ({\cal E}_m - \overline{E}_\alpha)^2 / (2
\Delta^2) \} \ .
\ea
Eq.~(\ref{s7}) was also obtained recently for a system of interacting
spins~\cite{Kea15}. We use Eq.~(\ref{s7}) in the main part of the
paper. In Section~\ref{ens} we show how the use of Eq.~(\ref{s8})
affects the time dependence of thermalization.

Collecting results we obtain
\ba
\label{s12}
      {\rm Tr} (A \rho(t)) &=& {\rm Tr} (A U(t) \Pi U^\dag(t)) \ .
\ea
In the HF basis, the time-evolution operator $U(t)$ has matrix elements
\ba
\label{s13}
U(t)_{m n} = \sum_\alpha O_{m \alpha} \exp \{ - i E_\alpha t / \hbar \}
O_{n \alpha} \ .
\ea
The eigenvalues $E_\alpha$ and the matrix elements $O_{m \alpha}$ are
random variables with distributions specified in and below
Eq.~(\ref{s5}). While $A$ and $\Pi$ are independent of time and
nonstatistical, $U(t)$ is time dependent and, via its dependence on
the matrix elements $O_{m \alpha}$ and eigenvalues $E_\alpha$, is
stochastic. These properties of $U(t)$ cause ${\rm Tr} (A \rho(t))$ to
be a time-dependent stochastic process.

It would not make sense to use the statistical assumptios on the
eigenvalues $E_\alpha$ and matrix elements $O_{m \alpha}$ in
Eq.~(\ref{s13}) right from the beginning, i.e., for all $t \geq 0$.
Starting at $t = 0$ from an arbitrary fixed basis of states $(\mu,
\nu)$, it takes a finite time $t_0$ for the Hamiltonian to build up
the correlations of eigenfunctions and eigenvalues and their
statistical properties defined around Eq.~(\ref{s5}). We are
interested here in the long-time development of ${\rm Tr} (\rho(t))$
and do not address the transients that occur in the time interval $0
\leq t \leq t_0$. We use the statistical assumptions~(\ref{s5}) in
Eqs.~(\ref{s12}, \ref{s13}) only after the transients have subsided,
i.e., for times $t > t_0$, without mentioning that restriction
hereafter.

To characterize the semiclassical regime of large energy $E$ and to
display explicitly the condition $\delta E \ll E$, we calculate mean
value and second moment of $H$. We identify the energy $E$ of the
system with $\big\langle {\rm Tr} (H \Pi) \big\rangle$. We use
Eqs.~(\ref{s3}, \ref{s5}, \ref{s7}) and find
\ba
\label{s9}
E = \big\langle {\rm Tr} (H \Pi) \big\rangle &=& \sum_m {\cal E}_m
\Pi_{m m} = {\rm Tr} (H_{\rm HF} \Pi) \ .
\ea
The energy $E$ is defined in terms of the HF Hamiltonian only. On
average the statistical fluctuations do not contribute to $E$. For $E$
to be in the semiclassical regime, the diagonal matrix elements of
$\Pi$ in the HF basis must be concentrated on a corresponding range of
large HF energies.

For the calculation of the second moment we assume that $\Delta$ is
constant (i.e., independent of energy) over the range of HF energies
where the matrix elements $\Pi_{m m}$ are essentially different from
zero and use ${\rm Tr} (\Pi) = 1$. We find
\ba
\label{s10}
\big\langle {\rm Tr} (H^2 \Pi) \big\rangle &=& \sum_m ({\cal E}^2_m +
\Delta^2)) \Pi_{m m} \nonumber \\
&=& {\rm Tr} (H^2_{\rm HF} \Pi) + \Delta^2 \ .
\ea
An estimate of the square of the energy spread $\delta E$ is given by
\ba
\label{s11}
(\delta E)^2 &\approx& \big\langle {\rm Tr} (H^2 \Pi) \big\rangle -
(\big\langle {\rm Tr} (H \Pi) \big\rangle)^2 \nonumber \\
&=& \bigg( {\rm Tr} (H^2_{\rm HF} \Pi) - [{\rm Tr} (H_{\rm HF} \Pi)]^2
\bigg) + \Delta^2 \ .
\ea
Eqs.~(\ref{s9}) to (\ref{s11}) show that for the inequality $\delta E
\ll E$ to hold, the range of HF states for which the diagonal elements
$\Pi_{m m}$ essentially differ from zero, must be sufficiently
narrow. The big round bracket in Eq.~(\ref{s11}) is positive or
zero. That implies $\delta E \geq \Delta$ for every choice of
$\Pi$. The big round bracket vanishes when in the HF basis $\Pi$ is
restricted to a single nonzero diagonal element so that $\Pi_{m m'} =
\delta_{m m_0} \delta_{m' m_0}$. Even then, the statistical
fluctuations of the random-matrix ensemble cause a minimum energy
spread of size $\Delta$. The condition $\delta E \ll E$ then reads
$\Delta \ll E$.

\section{Ensemble Average}
\label{ens}

In the present Section we calculate the ensemble average $\langle {\rm
  Tr} (A \rho(t)) \rangle$ of ${\rm Tr} (A \rho(t))$. As done in the
derivation of Eq.~(\ref{s10}), we assume that $\Delta$ is independent
of energy in the range of HF energies where the matrix elements of
$\Pi$ in the HF basis are essentially different from zero. Without
that assumption, the increased complexity of the notation would only
obscure the basic simplicity of our arguments and results, without
furnishing additional insights.

Using expression~(\ref{s13}) we calculate first the ensemble average
over the elements $O_{m \alpha}$ of the orthogonal matrix, keeping the
eigenvalues $E_\alpha$ fixed. The elements $O_{m \alpha}$ are
zero-centered Gaussian-distributed random variables with second
moments given in Eqs.~(\ref{s5}) and (\ref{s7}). We find
\ba
\label{e1}
&& \langle U(t)\rangle_{m n} = \frac{1}{\sqrt{2 \pi} \rho({\cal E}_m)
  \Delta} \delta_{m n} \nonumber \\ && \times \sum_\alpha \exp \{
({\cal E}_m - \overline{E}_\alpha)^2 / (2 \Delta^2) - i E_\alpha t /
\hbar \} \ .
\ea
Here and in Eq.~(\ref{e2}), angular brackets denote only the average
over the matrix elements. The variances are defined as
\ba
\label{e2}
v_1 &=& \langle U(t_1)_{m n} U^*(t_2)_{m' n'} \rangle - \langle
U(t_1) \rangle_{m n} \langle U^*(t_2) \rangle_{m' n'} \ , \nonumber \\ 
v_2 &=& \langle U(t_1)_{m n} U(t_2)_{m' n'} \rangle - \langle
U(t_1) \rangle_{m n} \langle U(t_2) \rangle_{m' n'} \ .
\ea
A term carrying four Gaussian-distributed matrix elents $O_{m \alpha}$
is averaged by ``contraction'', i.e., by applying Eqs.~(\ref{s5},
\ref{s8}) to all pairs of matrix elements, a total of four
possibilities. We find
\ba
\label{e3}
&& v_{1, 2} = \sum_\alpha \frac{1}{2 \pi \rho^2({\cal E}_m) \Delta^2}
(\delta_{m n'} \delta_{n m'} + \delta_{m m'} \delta_{n n'}) \nonumber \\
&& \ \ \times \exp \{ - ({\cal E}_m - \overline{E}_\alpha)^2 / ( 2 \Delta^2)
- ({\cal E}_n - \overline{E}_\alpha)^2) ( 2 \Delta^2) \} \nonumber \\
&& \ \ \times \exp \{ i E_\alpha (t_2 \mp t_1) / \hbar \} \ . 
\ea
The upper (lower) sign holds, respectively, for $v_1$ (for
$v_2$). Using Eq.~(\ref{e1}), its complex conjugate, and the
result~(\ref{e3}) for $v_1$ in Eq.~(\ref{s12}) gives
\ba
\label{e4}
&& \big\langle {\rm Tr} (A \rho(t)) \big\rangle = \sum_{m n}
\frac{\exp \{ i ({\cal E}_m - {\cal E}_n) t / \hbar \} }{2 \pi
  \rho({\cal E}_m) \rho({\cal E}_n) \Delta^2}  A_{m n} \Pi_{n m}
\nonumber \\
&& \ \ \times \bigg\langle \sum_\alpha \exp \{ - ({\cal E}_m -
\overline{E}_\alpha)^2 / (2 \Delta^2) + i E_\alpha t / \hbar \} \\
&& \ \ \times \sum_\beta \exp \{ - ({\cal E}_n - \overline{E}_\beta)^2 /
(2 \Delta^2) - i E_\beta t / \hbar \} \bigg\rangle \nonumber \\
&& + \sum_{m n} \frac{1}{2 \pi \rho^2({\cal E}_m) \Delta^2}
(A_{m m} \Pi_{n n} + A_{n m} \Pi_{n m}) \nonumber \\
&& \times \sum_\alpha \exp \{ - [({\cal E}_m - \overline{E}_\alpha)^2 +
  ({\cal E}_n - \overline{E}_\alpha)^2)] / ( 2 \Delta^2) \} \ .
\nonumber
\ea
The big angular brackets denote the remaining average over the
eigenvalues. The average consists of two contributions resulting,
respectively, from averaging the sums over $\alpha$ and $\beta$
separately, and from the correlation function of $E_\alpha$ and
$E_\beta$. The latter is given by the two-point GOE eigenvalue
correlation function~\cite{Meh04}. That is a function of $(E_\alpha -
E_\beta) \rho(E_\alpha)$ which differs significantly from zero only
over the range of a few mean level spacings. Therefore, the correlated
part of the average in Eq.~(\ref{e4}) is small of order $1 /
N_\Delta$. That is shown explicitly in Ref.~\cite{Wei23}. We keep the
uncorrelated part only, i.e., we perform the sums over $\alpha$ and
$\beta$ independently. In the sum over $\alpha$ we write $({\cal E}_m
- \overline{E}_\alpha) = ({\cal E}_m - E_\alpha) + (E_\alpha -
\overline{E}_\alpha)$. While $({\cal E}_m - E_\alpha)$ extends over
the range $\Delta$, the fluctuations $(E_\alpha -
\overline{E}_\alpha)$ are confined to a few mean level
spacings. Therefore, the term $(E_\alpha - \overline{E}_\alpha)$ is
small of order $1 / N_\Delta$ compared to $({\cal E}_m - E_\alpha)$,
and is neglected. We proceed analogously for the sum over $\beta$. All
sums over $\alpha$ and $\beta$ in Eq.~(\ref{e4}) are then carried out
in the continuum limit. We find
\ba
\label{e5}
&& \big\langle {\rm Tr} (A \rho(t)) \big\rangle = {\rm Tr} (A \Pi)
\exp \{ - t^2 \Delta^2 / \hbar^2 \} \nonumber \\
&& \ \ + \sum_{m n} \frac{1}{2 \sqrt{\pi} \rho({\cal E}_m) \Delta}
(A_{m m} \Pi_{n n} + A_{n m} \Pi_{n m}) \nonumber \\
&& \ \ \ \ \times \exp \{ - ({\cal E}_m - {\cal E}_n)^2 / ( 4 \Delta^2)
\} \ .
\ea
We simplify the terms on the right-hand side of Eq.~(\ref{e5}),
neglecting terms of order $1 / N_\Delta$. We use the eigenbasis of the
operator $\Pi$ with orthonormal eigenvectors $\langle n | \kappa
\rangle$ and with eigenvalues $\pi_\kappa$ that obey $\sum_\kappa
\pi_\kappa = 1$ and $0 \leq \pi_\kappa \leq 1$ for all $\kappa$. Then
${\rm Tr} (A \Pi) = \sum_\kappa A_{\kappa \kappa} \pi_\kappa$. The
constraints on the eigenvalues $\pi_\kappa$ imply that ${\rm Tr} (A
\Pi)$ is of order unity. We show that the same is true for the term
proportional to $A_{m m} \Pi_{n n}$ while the term proportional to
$A_{m n} \Pi_{n m}$ will be shown to be of order $1 / N_\Delta$. The
term proportional to $A_{m m} \Pi_{n n}$ carries a double sum over $m$
and $n$. For our estimate we divide the summation over $m$ into energy
intervals of length $\Delta$ each, labeled consecutively by $k = 1, 2,
\ldots$. For fixed $m \in k$, the sum over $n$ can be read as the
partial trace of $\Pi$ extended over the HF states in a neighborhood
of width $\sqrt{2} \Delta$ of ${\cal E}_m$. For all states $m \in k$
we approximate that partial trace of $\Pi$ by an expression $p_k$
that is independent of $m$ and given by
\ba
\label{e6}
p_k = \sum_{n \in k} \sum_\kappa |\langle n | \kappa
  \rangle|^2 \pi_\kappa \ .
\ea
Here $0 \leq p_k \leq 1$, and $\sum_k p_k = 1$. The
approximation~(\ref{e6}) may be off by a numerical factor but not by a
factor of order $N_\Delta$. The term proportional to $A_{m m} \Pi_{n
  n}$ takes the form $\sum_k p_k [{\rm Tr}_k (A)] / (\sqrt{2} \pi
\rho_k \Delta)$. Here $\rho_k$ is the average level density in the
interval labeled $k$, and ${\rm Tr}_k$ denotes the trace taken over
the HF states in that interval. Since ${\rm Tr}_k (A)$ is of order
$N_\Delta$, the term is of order unity. Using similar arguments
we show in the Appendix that the last term on the right-hand side of
Eq.~(\ref{e5}) is proportional to $1 / N_\Delta$. That term is
neglected.

In summary, we have shown that for $1 / N_\Delta \ll 1$, the ensemble
average is given by
\ba
\label{e7}
&& \big\langle {\rm Tr} (A \rho(t)) \big\rangle \nonumber \\ && = {\rm
  Tr} (A \exp \{ - i H_{\rm HF} t / \hbar \} \Pi \exp \{ + i H_{\rm
  HF} t / \hbar \}) \nonumber \\
&& \ \ \ \ \times \exp \{ - t^2 \Delta^2 / \hbar^2 \} \nonumber \\
&& + \sum_k \frac{p_k}{\sqrt{2} \pi \rho_k \Delta} {\rm Tr}_k (A) \ .
\ea
Eq.~(\ref{e7}) rests on the assumption~(\ref{e6}). That assumption is
not absolutely necessary. For an accurate treatment it is possible to
retain the original form of the term proportional to $A_{m m} \Pi_{n
  n}$. However, the asssumption~(\ref{e6}) simplifies that term and
makes the result physically more transparent. We keep the
form~(\ref{e7}) for the remainder of the paper.

The central approximation used in deriving Eq.~(\ref{e7}) is $N_\Delta
\gg 1$. As shown in Section~\ref{hart}, that condition confines the
statistical operator to high energies. That is consistent with the
condition ``energy $E$ in the semiclassical regime'' stated in the
literature~\cite{Sre99, DAl16, Aba19} as necessary for thermalization.

We discuss the result~(\ref{e7}). The first term on the right-hand
side describes the relaxation of the term $\langle {\rm Tr} (A
\rho(t)) \rangle$ as function of time $t$. Starting at time $t = 0$
with ${\rm Tr} (A \Pi)$, the term vanishes for $t \to \infty$. The
governing factor for relaxation is the Gaussian with characteristic
time scale $\Delta$. The Gaussian is the square of the Fourier
transform of the correlation function $F(x)$ in Eq.~(\ref{s7}). That
is consistent with the result of Ref.~\cite{Wei23} where the time
dependence of relaxation was found to be given by the square of the
Fourier transform of the average GOE level density $\sqrt{1 - E^2 / (
  4 \lambda^2)}$. It is easily verified that use of the
Lorentzian~(\ref{s8}) for $F(x)$ results in an exponential factor
$\exp \{ - \Delta t \}$ instead of the Gaussian. We conclude that the
correlation width $\Delta$ determines the time scale for relaxation
irrespective of the form of $F(x)$. The actual analytical form of the
relaxation function depends upon the choice of the function $F(x)$.

For time $t \to \infty$, $\langle {\rm Tr} (A \rho(t)) \rangle$
approaches asymptotically the last term in Eq.~(\ref{e7}). If the
system thermalizes, that term must be equal to ${\rm Tr} (A \rho_{\rm
  eq})$. For an unrestricted statistical operator $\Pi$, that is, in
general, not the case. The statistical operator may be spread over
several or many intervals $k$. Then a corresponding number of the
coefficients $p_k$ in Eq.~(\ref{e7}) have values that differ
significantly from zero. For an arbitrary distribution of the
coefficients $p_k$ the asymptotic value of $\langle {\rm Tr} (A
\rho(t)) \rangle$ differs from the equilibrium value ${\rm Tr} (A
\rho_{\rm eq})$. Thermalization occurs only if the statistical
operator $\Pi$ is confined to a single interval $k_0$ or, less
strictly, if $\sum_{m, n \in k_0} |\Pi_{m n}|^2 \gg \sum_{m, n \in k}
|\Pi_{m n}|^2$ for all $k \neq k_o$ and if $\sum_{m, n \in k_0}
|\Pi_{m n}|^2 \gg \sum_{m \in k_1, n \in k_2} |\Pi_{m n}|^2$ for all
pairs of intervals $(k_1 \neq k_2)$. Then the approximation~(\ref{e6})
becomes nearly exact, we have $p_k \approx \delta_{k k_0}$, and the
last term in Eq.~(\ref{e7}) is given by $(1 / N_{k_0}) {\rm Tr}_{k_0}
(A)$. That is equal to the equilibrium value ${\rm Tr} (A \rho_{\rm
  eq})$ because, with $\Pi$ confined to the interval $k_0$,
equilibrium corresponds to equal occupation probability of all states
in $k_0$. (For the narrow energy interval $\Delta$, the Boltzmann
factor is practically constant). We conclude that for thermalization
to occur the condition $\delta E \approx \Delta$ must hold. That
quantifies the usual condition $\delta E \ll E$ stated in the
literature~\cite{DAl16, Aba19}. Thermalization would also be attained
if several or many of the coefficients $p_k$ differ from zero and if
their distribution corresponds to statistical equilibrium. But that is
the case only if $\Pi$ itself is close to equilibrium, whereas
thermalization as formulated in expression~(\ref{i1}) is supposed to
hold in general, i.e., for every form of the statistical operator
subject only to the conditions ``large $E$'' and $\delta E \ll E$.

\section{Correlation Function}
\label{cor}

The stochastic process ${\rm Tr} (A \rho(t))$ is real because the
operators $A, H, \Pi$ are Hermitean. Therefore, the correlation
function is
\ba
\label{c1}
\big\langle {\rm Tr} (A \rho(t_1)) {\rm Tr} (A \rho(t_2)) \big\rangle
- \big\langle {\rm Tr} (A \rho(t_1)) \big\rangle \big\langle {\rm Tr}
(A \rho(t_2)) \big\rangle \ .
\ea
In calculating expression~(\ref{c1}) we order the terms by the number
$\nu$ of pairs of matrix elements $O_{m \alpha}$ that are
``cross-contracted'', with one element of the pair located in one
trace and the other element in the other. The elements $O_{m \alpha}$
are Gaussian random variables. Nonvanishing contributions may,
therefore, arise only for $\nu = 0, 2, 4$. Each term in
xpression~(\ref{c1}) contains two factors $U(t)$ and two factors
$U^\dag(t)$. Contraction of the matrix elements $O_{m \alpha}$ causes
the appearance of $\langle U(t) \rangle$, of $\langle U^\dag(t)
\rangle$, of $v_1$ and $v_2$ defined in Eqs.~(\ref{e2}), and of the
third- and fourth-order cumulants of $U(t)$. Our results are written
in terms of these cumulants. As in Section~\ref{ens} we use a division
of the energy scale into intervals of equal length $\Delta$ that are
consecutively labeled $k = 1, 2, \ldots$.

For $\nu = 0$, all four matrix elements $O_{m \alpha}$ in either trace
are contracted with each other, and nonzero contributions to the
correlation function~(\ref{c1}) arise only from the correlations of
the eigenvalues $E_\alpha$. That situation is realized if in either
trace both factors $U$ and $U^\dag$ are replaced by their mean
values~(\ref{e1}) or by their variances $v_1$ in Eqs.~(\ref{e3}),
resulting in a total of four possibilities. When all four factors $U,
U^\dag$ are replaced by their mean values, the Gausssian factors and
the Kronecker deltas in Eq.~(\ref{e1}) confine all summation indices
in either trace to the same intervals labeled $k_1$ and $k_2$,
respectively. We find
\ba
\label{c2}
&& \sum_{k_1, k_2} \frac{{\rm Tr}_{k_1} (A \Pi) {\rm Tr}_{k_2} (A \Pi)} 
      {(\sqrt{2 \pi} \rho_{k_1} \Delta)^2 (\sqrt{2 \pi} \rho_{k_2}
        \Delta)^2} \sum_{\alpha, \beta \in k_1}
      \sum_{\gamma, \delta \in k_2} \nonumber \\
      && \times \bigg\langle \exp \{ i (E_\alpha - E_\beta) t_1 / \hbar
      + i (E_\gamma - E_\delta) t_2 / \hbar \} \bigg\rangle_{\rm corr} \ . 
\ea
The index ``corr'' indicates the correlated part. The uncorrelated
part would give the product of two factors, each of the form of the
first term on the right-hand side of Eq.~(\ref{e5}), taken,
respectively, at times $t_1$ and $t_2$. Both these factors are of
order unity in $N_{k_1}, N_{k_2}$. Correlations between the
eigenvalues $E_\alpha, E_\beta \in k_1$ and $E_\gamma, E_\delta \in
k_2$ vanish unless $k_1 = k_2$. Within the interval $k_1 = k_2$
contributions to the big angular bracket arise from the GOE two-point
correlation function~\cite{Meh04} of pairs of eigenvalues $(E_\alpha,
E_\gamma)$, $(E_\alpha, E_\delta)$, $(E_\beta, E_\gamma)$, or
$(E_\beta, E_\delta)$, from the product of two such correlation
functions, and from the GOE four-point function. As mentioned in
Section~\ref{ens}, the GOE two-point correlation function is small of
order $1 / N_{k_1}$. The four-point function is even
smaller~\cite{Meh04}. Therefore, the contribution~(\ref{c2}) to the
correlation function~(\ref{c1}) is negligible. The same argument shows
that the remaining three of the four possibilities mentioned above
expression~(\ref{c2}) are likewise negligible.

For $\nu = 2$ and for $\nu = 4$, cross-contraction of pairs of matrix
elements $O_{m \alpha}$ does give nonvanishing contributions. In each
of these, the leading-order term in $1 / N_\Delta$ arises when
correlations between eigenvalues $E_\alpha, E_\beta, E_\gamma,
E_\delta$ are neglected. Therefore, the summations over these
eigenvalues may be carried out in the expressions for $\langle U(t)
\rangle$ and in the higher-order cumulants of $U$ prior to using these
in the correlation function~(\ref{c1}). We do that using the
approximation introduced below Eq.~(\ref{e4}). For $\langle U(t)
\rangle$ in Eq.~(\ref{e1}) and for $v_1, v_2$ in Eqs.~(\ref{e3}) that
gives
\ba
\label{c3}
\langle U(t)_{m n} \rangle &=& \delta_{m n} \exp \{ - t^2 \Delta^2 /
(2 \hbar^2) \} \ , \nonumber \\
v_1 &=& \frac{1}{2 \sqrt{\pi} \rho({\cal E}_m) \Delta} (\delta_{m n'}
\delta_{n m'} + \delta_{m m'} \delta_{n n'}) \nonumber \\
&& \qquad \times \exp \{ - ({\cal E}_m - {\cal E}_n)^2 / (4 \Delta^2)
\} \ , \nonumber \\
v_2 &=& \frac{1}{2 \sqrt{\pi} \rho({\cal E}_m) \Delta} (\delta_{m n'}
\delta_{n m'} + \delta_{m m'} \delta_{n n'}) \nonumber \\
&& \qquad \times \exp \{ - ({\cal E}_m - {\cal E}_n)^2 / (4 \Delta^2)
\} \nonumber \\
&& \qquad \times \exp \{ + i (t_1 + t_2) ({\cal E}_m + {\cal E}_n) /
\hbar \} \nonumber \\
&& \qquad \times \exp \{ - (t_1 + t_2)^2 \Delta^2 / (4 \hbar^2) \} \ .
\ea
While the index $m$ in the expression for $\langle U(t)_{m n} \rangle$
is unrestricted, the indices $(m, n)$ in the expressions for $v_1$ and
$v_2$ are confined to the lie in the same interval of width
$\Delta$. In addition to the terms in Eq.~(\ref{c3}),
cross-contraction of the matrix elements $O_{m \alpha}$ also gives rise
to the appearance of the third- and fourth-order cumulants of $U(t)$.
In the Appendix we show that contributions of these higher-order
cumulants to the correlation function~(\ref{c1}) are negligible. In the
present Section, we confine ourselves to the contributions~(\ref{c3}).

For $\nu = 2$, there are four possibilities. One factor $U$ or
$U^\dag$ in either trace is replaced by its average, the remaining two
factors $U$ or $U^\dag$ being replaced by $v_1$ or $v_2$ as the case
may be. As a representative example we consider the replacements
\ba
\label{c4}
&& U^\dag(t_1) \to \langle U^\dag(t_1) \rangle \ , \ U(t_2) \to \langle
U(t_2) \rangle \ , \nonumber \\
&& U(t_1) U^\dag(t_2) \to v_1 \ .
\ea
Using Eqs.~(\ref{c3}) we obtain
\ba
\label{c5}
&& \sum_k \frac{1}{2 \pi \rho_k \Delta} \sum_{m, n \in k} [\Pi A]_{m n}
\big( [A \Pi]_{n m} + [A \Pi]_{m n} \big) \nonumber \\
&& \qquad \times \exp \{ - (t_1^2 + t_2^2) \Delta^2 / (2 \hbar^2) \} \ . 
\ea
Here and in the remainder of the Section, the symbol $[A \Pi]_{m n} =
\sum_l A_{m l} \Pi_{l n}$ stands for the matrix product $A \Pi$ with
unrestricted summation over intermediate states $l$. Without
additional assumptions, the term~(\ref{c5}) cannot be shown to be
small of order $1 / N_\Delta$. The term~(\ref{c5}) has the same time
dependence as the first term on the right-hand side of
Eq.~(\ref{e7}). It gives the statistical fluctuations in the ensemble
of matrices $H$ around that term. Analogous results are obtained for
the other three possibilities beyond the choice~(\ref{c4}).

For $\nu = 4$, all four factors $U, U^\dag$ are replaced either by two
factors $v_1$ or by two factors $v_2$, the factors $U$ or $U^\dag$
contributing to $v_1$ or $v_2$ being taken from either trace. A
representative example is given by the replacements
\ba
\label{c6}
U(t_1) U^\dag(t_2) \to v_1 \ , \ U^\dag(t_1) U(t_2) \to v_1
\ea
which give
\ba
\label{c7}
&& \sum_{k_1, k_2} \frac{1}{ (2 \pi)^2 \rho_{k_1} \rho_{k_2} \Delta^2}
\sum_{n, p \in k_1} \sum_{s, m \in k_2} A_{m n} \Pi_{p s} \\
&& \times \big( A_{p s} \Pi_{m n} + A_{p m} \Pi_{s n} + A_{n s} \Pi_{m p}
+ A_{n m} \Pi_{s p} \big) \ . \nonumber 
\ea
Every matrix element of the operators $A$ and $\Pi$ appearing in
expression~(\ref{c7}) connects states in the two intervals $(k_1,
k_2)$. The second and third term in round brackets can be written as a
single trace in the interval $k_1$. These terms are of order $N_{k_1}$
and, multiplied by $1 / [(2 \pi)^2 \rho_{k_1} \rho_{k_2} \Delta^2]$,
are neglibible. The last term is equal to $(\sum_{m n} A_{m n} A_{n
  m}) (\sum_{p s} \Pi_{p s} \Pi_{s p})$. The first factor is of order
$N_{k_1}$. The second factor is of order unity. To see that we use
$\Pi = \Pi^\dag$ and write the factor as $\sum_{p s} | \Pi_{p s} |^2$.
That expression is bounded from above by ${\rm Tr}_{k_1} (\Pi^2) < 1$.
Together with the factor $ 1 / [(2 \pi)^2 \rho_{k_1} \rho_{k_2}
  \Delta^2]$ the last term is negligible. That leaves us with
\ba
\label{c8}
&& \sum_{k_1, k_2} \frac{1}{ (2 \pi)^2 \rho_{k_1} \rho_{k_2} \Delta^2}
\bigg( \sum_{n \in k_1, m \in k_2} (A_{m n} \Pi_{m n}) \bigg)^2 \ .
\ea
Without additional assumptions, the contribution~(\ref{c8}) to the
correlation function~(\ref{c1}) cannot be shown to be negligible. The
term~(\ref{c8}) is independent of time and describes the statistical
fluctuations of the ensemle of Hamiltonians $H$ around the last term
in Eq.~(\ref{e5}).

Both terms~(\ref{c5}) and (\ref{c8}) are negligible, however, if
either the operator $A$ or the operator $\Pi$ are essentially diagonal
in HF space. That condition says that for all intervals $k_1 \neq k_2$
we have $\sum_{m, n \in k_1} |A_{m n}|^2 \gg \sum_{n \in k_1, m \in
  k_2} |A_{m n}|^2$, and correspondingly for $\Pi$. For $A$, that
condition is similar to a condition formulated in Ref.~\cite{Wei23}
and to the ``eigenvalue thermalization hypothesis'' of
Ref.~\cite{Sre99}. The condition for $\Pi$ coincides with the
condition for thermalization formulated in Section~\ref{ens}. We
conclude that if the system thermalizes, the statistical fluctuations
around the terms in Eq.~(\ref{e7}) are vanishingly small. That
statement holds without constraining assumptions on the operator
$A$. Then all systems described by the stochastic process ${\rm Tr} (A
\rho(t))$ thermalize in the same manner, including the system
described by the Hamiltonian defined in the first part of
Section~\ref{stat}.

\section{Violation of Time-Reversal Invariance}
\label{vio}

With proper modifications, the arguments used in Section~\ref{stat}
apply also to the case of time-reversal non-invariant systems, and the
results in Sections~\ref{ens} and \ref{cor} hold as well. The BGS
conjecture now postulates local agreement of the spectral fluctuation
properties of the Hamiltonian $H$ with those of the Gaussian Unitary
Ensemble (GUE) of random matrices~\cite{Meh04}. Proceeding as in
Section~\ref{stat}, we replace Eq.~(\ref{s3}) by
\ba
\label{v1}
H_{m n} = \sum_\alpha {\cal U}_{m \alpha} E_\alpha \ {\cal U}^*_{n \alpha} \ . 
\ea
The matrices ${\cal U}$ are unitary. Both the eigenvalues $E_\alpha$
and the eigenfunctions ${\cal U}_{m \alpha}$ obey local GUE
statistics. Eigenvalues and eigenfunctions are uncorrelated. In an
interval of width $\Delta$, the eigenvalues follow the level
statistics of the GUE, and the eigenfunctions ${\cal U}_{m \alpha}$
are zero-centered complex Gaussian random variables with second
moments
\ba
\label{v2}
\langle {\cal U}_{m \alpha} \ {\cal U}_{n \beta} \rangle &=& 0 \ ,
\nonumber \\ \big\langle {\cal U}_{m \alpha} \ {\cal U}^*_{m' \alpha'}
\big\rangle &=& \delta_{m m'} \delta_{\alpha \alpha'} F[({\cal E}_m -
  \overline{E}_\alpha)^2 / \Delta^2] \ .
\ea
The function $F(x)$ is given by Eq.~(\ref{s7}). The contraction rules
used in calculating the ensemble average and the correlation function
of ${\rm Tr} (A \rho(t))$ are changed. In the orthogonal case, every
matrix element $O_{m \alpha}$ is contracted with every other such
matrix element. Because of Eqs.~(\ref{v2}), only contractions of
${\cal U}$ with ${\cal U}^\dag$ occur in the present case. That
reduces the number of contraction terms. For instance, the term $v_2$
in Eqs.~(\ref{e2}) does not arise, and the term proportional to $A_{n
  m} \Pi_{n m}$ in Eq.~(\ref{e5}) is lacking. The result~(\ref{e7})
remains unchanged, however. The absence of a number of terms does not
affect the structure of the results in Section~\ref{cor}. The
arguments used and conclusions drawn there remain the same. We
conclude that relaxation and thermalization are completely similar for
systems that are and systems that are not, invariant under time
reversal.

\section{Discussion and Summary}
\label{dis}

We have investigated the thermalization of a closed chaotic many-body
quantum system by combining the Hartree-Fock (HF) approach with the
Bohigas-Giannoni-Schmit (BGS) conjecture. The HF approach provides the
scaffolding for the spectrum of the many-body Hamiltonian $H$. It
defines the integrable Hartree-Fock Hamiltonian $H_{\rm HF}$, the
average HF level density $\rho(E)$ as a function of excitation energy,
and it allows expectation values of physical operators to take
different values in different parts of the spectrum. In particular, it
allows for a physically meaningful definition of the semiclassical
regime. We assume that the total Hamiltonian, given by $H = H_{\rm HF}
+ V$, describes a chaotic quantum system. We accordingly use the BGS
conjecture. We consider the conjecture to be universally valid and to
be synonymous with quantum chaos. The conjecture states that locally,
the spectral fluctuation properties of $H$ coincide with those of the
random-matrix ensemble in the same symmetry class. The word
``locally'' means that for every energy $E$ of the system, the
coincidence is restricted to an energy interval of width $\Delta$ (the
correlation width) centered on $E$. In the framework of the HF
approach, the BGS conjecture requires that within the interval
$\Delta$, the mixing of HF states due to the residual interaction $V$
is so strong that the resulting eigenstates of $H$ approach
characteristic statistical properties of GOE eigenstates. Such strong
local mixing of the HF states requires $V$ to be sufficiently
strong. At the same time we have assumed that the overall structure of
the spectrum and, in particular, the average level density $\rho(E)$,
are not affected by $V$. Pictorially speaking, $V$ puts a thin veneer
on the scaffold of the spectrum provided by the HF approach.

Combining the HF approach and the BGS conjecture we, thus, postulate
that locally the statistical properties of eigenfunctions and
eigenvalues of $H$ coincide with those of the GOE in an interval
defined by a Gaussian of width $\Delta$. Actually, the statistical
properties of the GOE that we use hold in the limit of infinite matrix
dimension. The required coincidence (and, therefore, the validity of
the BGS conjecture) depends on the number $N_\Delta = \Delta \rho(E)$
of states in the interval $\Delta$ and grows with $N_\Delta$. Using
essentially Fermi's golden rule, we have estimated the correlation
width $\Delta$. We have shown that $\Delta$ depends on energy $E$ much
less strongly than $\rho(E)$. Since $\rho(E)$ grows essentially
exponentially with $E$, the condition $N_\Delta \gg 1$ is best
fulfilled in the semiclassical regime. In that regime, we may neglect
terms of order $1 / N_\Delta$.

Using the statistical properties of $H$ in the evaluation of ${\rm Tr}
(A \rho(t))$ poses a challenge. We solve the problem with the help of
the following construction. Changing the eigenvalues and
eigenfunctions of $H$ into random variables with GOE properties, we
define an ensemble of Hamiltonians, all with the same statistical
properties as the original Hamiltonian $H$. Written in terms of that
ensemble, the time-dependent function ${\rm Tr} (A \rho(t))$ turns
into a stochastic process. We calculate statistical average and
correlation function of that process in the limit $N_\Delta \gg
1$. The average $\langle {\rm Tr} (A \rho(t)) \rangle$ is given by our
central result Eq.~(\ref{e7}). Vanishing of the correlation
function~(\ref{c1}) guarantees that Eq.~(\ref{e7}) holds for all
members of the ensemble and, thus, for our original Hamiltonian. The
condition for that to happen is addressed below.

Eq.~(\ref{e7}) shows that $\langle {\rm Tr} (A \rho(t)) \rangle$
relaxes in time with a Gaussian factor of width $\Delta$. That factor
is due to the Gaussian form of the correlation
function~(\ref{s7}). Other forms lead to different time-dependent
functions. For example, the Lorentzian in Eq.~(\ref{s8}) yields an
exponential. All these forms have in common that the time scale $\hbar
/ \Delta$ is set by the correlation width $\Delta$.

For time $t \to \infty$, the right-hand side of Eq.~(\ref{e7}) relaxes
to the time-independent term. That term depends upon the operator $A$
and on the time-independent statistical operator $\Pi$ which defines
the distribution of the system over the states in Hilbert space. The
operator $\Pi$ determines whether or not $\langle {\rm Tr} (A \rho(t))
\rangle$ thermalizes. For a general form of $\Pi$, the coefficients
$p_k$ in Eq.~(\ref{e7}) may take any value, and $\langle {\rm Tr} (A
\rho(t)) \rangle$ differs from ${\rm Tr} (A \rho_{\rm eq})$. Moreover,
evaluation of the correlation function in Section~\ref{cor} shows
that in that case there exist statistical fluctuations (both
time-independent and time dependent and of Gaussian form) around the
mean value~(\ref{e7}).

Thermalization occurs only if constraining conditions are imposed on
$\Pi$. The standard requirement~\cite{DAl16, Aba19} is that the
average energy $E$ of the system be in the semiclassical regime, and
that the spread $\delta E$ in energy around that average be small. In
the present context, the first condition requires that the diagonal
elements of $\Pi$ in the HF basis essentially differ from zero only in
the domain of large HF energies. That validates the inequality
$N_\Delta \gg 1$. The spread $\delta E$ in energy, expressed in terms
of $\Pi$, is estimated as the root-mean-square deviation of the energy
from its average value $E$. The result shows that $\delta E \geq
\Delta$ so that $\Delta$ is the minimum energy spread in the ensemble
of random Hamiltonians. Thermalization is seen to occur universally
(irrespective of the detailed form of $\Pi$) only if $\Pi$ is confined
to HF states within an energy interval of width $\Delta$. Then the
last term of Eq.~(\ref{e7}) is equal to ${\rm Tr} (A \rho_{\rm
  eq})$. If that condition is met, the statistical fluctuations around
the mean value $\langle {\rm Tr} (A \rho(t)) \rangle$ vanish. All
members of the stochastic process ${\rm Tr} (A \rho(t))$ (including
the one containing our original Hamiltonian) tend asymptotically in
time towards the equilibrium value ${\rm Tr} (A \rho_{\rm eq})$. These
conclusions hold both for systems that are and for systems that are
not invariant under time reversal.

Our approach leaves open several questions. We have assumed that the
average level density of the full Hamiltonian is the same as that of
the HF Hamiltonian. That puts an upper bound on the strength of $V$
which we have not identified. We have not investigated collective
effects. In atomic nuclei, for instance, the presence of collective
states strongly modifies the average level density at low energy, see,
for instance, Ref.~\cite{Gut21}. Does such an effect persist in the
semiclassical regime?  How do scars in the spectrum~\cite{Kap99}
affect our approach? We have assumed that two energy intervals (the
one that limits agreement with the $\Delta_3$ statistics of
eigenvalues and the one that limits complete mixing of eigenfunctions)
are equal. That assumption requires further scrutiny. We have assumed
that the spectral correlation function $F(x)$ is Gaussian. Other forms
have been considered in the literature. Which form applies for a given
system? The answer determines the time dependence of relaxation.

Our approach differs from, and our results go beyond those of
Refs.~\cite{Sre99, DAl16, Aba19}. We deduce thermalization from
constraints on the operator $\Pi$ and from the statistical properties
of the Hamiltonian. The latter, in turn, follow stringently from the
BGS conjecture. The conjecture holds within the correlation width
$\Delta$. That parameter plays the central role in our approach. It
determines the characteristic time scale for relaxation. It also
determines the value of the energy spread $\delta E$ required for
thermalization. The parameter $\Delta$ plays no role in
Refs.~\cite{Sre99, DAl16, Aba19}. Our approach makes a prediction that
can, in principle, be tested experimentally: The correlation width
$\Delta$, defined by the spectral properties of the Hamiltonian,
determines the time scale $\hbar / \Delta$ for relaxation of ${\rm Tr}
(A \rho(E))$.

We believe that our approach is not confined to chaotic many-body
systems but applies to chaotic quantum systems in general. Certainly,
the BGS conjecture holds, and the limiting interval $\Delta$ occurs,
generically, cf. the examples in Section~\ref{app}. The possibility to
construct an integrable Hamiltonian that defines the scaffolding of
the spectrum is, likewise, probably generic. Consider, for instance,
the quantum Sinai billard, a square with an inscribed concentric
circle for which the eigenfunctions of the Laplacian operator in two
dimensions vanish on both surfaces. Replacing the circle by a polygon
with $n$ corners and the same area as the circle one obtains an
integrable system the spectrum of which for $n \gg 1$ provides the
scaffolding for the spectrum of the Sinai billard. The geometric
difference between polygon and circle is the analogue of the residual
interaction $V$ in the present paper, mixing close-lying eigenstates
of the polygon within the interval $\Delta$. If generic, the
possibility to find for every chaotic quantum system an integrable
Hamiltonian that differs from the chaotic one only by a weak residual
interaction, would qualitatively explain why the mixing of the
eigenstates of the integrable Hamiltonian is local and confined to the
correlation width $\Delta$.

The correlation width $\Delta$ is present universally in all chaotic
quantum systems that obey the BGS conjecture. We are not aware of any
other such common energy scale that would qualify for defining the
time scale for relaxation. That raises the question whether $\Delta$
possesses a classical counterpart that is present equally universally
in chaotic classical systems. In Ref.~\cite{Arv81}, the upbend of the
$\Delta_3$ statistics for the quantum Sinai billard was related to the
largest Ljapunov coefficient of the classical Sinai billard. A bound
on the agreement of the $\Delta_3$ statistics with the GOE prediction
was also discussed in the context of periodic-orbit
theory~\cite{Ber85}. We believe that this is an interesting question
that deserves further investigation.

\section*{Appendix: Estimates}

We estimate the term proportional to $A_{n m} \Pi_{n m}$ in
Eq.~(\ref{e5}). We use the intervals labeled $k$ introduced in
Section~\ref{ens}. In each such interval the summations over $(m, n)$
are both restricted to an energy interval of width $\Delta$. Assuming
that the sums are sufficiently smooth we approximate the term by
$\sum_k \sum_{m \in k, n \in k} A_{n m} \Pi_{n m} / (2 \sqrt{\pi}
\rho({\cal E}_m) \Delta)$. For each $k$, the operator $\Pi_{n m} =
\sum_\kappa \langle n | \kappa \rangle \pi_\kappa \langle \kappa | m
\rangle$ with $(m \in k, n \in k)$ is positive semidefinite and has
trace $p_m$. The eigenvalues $\pi_{m, \kappa}$ of that restricted
operator differ from the eigenvalues $\pi_\kappa$ of $\Pi$ and obey $0
\leq \pi_{m, \kappa} \leq p_m$ and $\sum_\kappa \pi_{m, \kappa} =
p_m$. In the eigenbasis we have $\sum_{m \in k, n \in k} A_{n m}
\Pi_{n m} = \sum_{\kappa \in m} A_{\kappa \kappa} \pi_{m,
  \kappa}$. That sum is of order unity. The claim is obvious if only a
single eigenvalue $\pi_{m \kappa}$ differs from zero, and if all
eigenvalues are equal and given by $1 / (\rho({\cal E}_m) \Delta)$. It
is easily seen that the constraints $\sum_\kappa \pi_{m \kappa} = 1$,
$0 \leq \pi_{m \kappa} \leq p_m$ lead to the same conclusion for other
choices of the eigenvalues $\pi_{m \kappa}$. We conclude that $\sum_{m
  \in k, n \in k} A_{n m} \Pi_{n m} / (2 \sqrt{\pi} \rho({\cal E}_m)
\Delta)$ is of order $1 / (\rho({\cal E}_m) \Delta) \ll 1$, and that
the term proportional to $A_{n m} \Pi_{n m}$ is negligible. Our
smoothness argument may be off by a numerical factor but that does not
invalidate the identification of the term as being small of order $1 /
(\rho \Delta)$.

We turn to the contribution of the third-order cumulant of $U$ to the
correlation function~(\ref{c1}). The form of that expression can be
obtained without detailed calculation. Contraction of the six matrix
elements $O_{m \alpha}$ in the third-order cumulant of $U$ yields
three Gaussian factors and three factors $1 / (\rho \Delta)$. The
three energies in the exponents of the three factors $U$ all carry the
same label $\alpha$ because the contractions are completely
linked. Summation over $\alpha$ generates a time-dependent Gaussian
factor, removes one factor $1 / (\rho \Delta)$, and confines the
remaining summations over $(m, n, p)$ to the same interval $k$. We are
left with the three operators $A, \Pi, [\Pi A]$. (The operator $[\Pi
  A]$ arises from the replacement $U^\dag(t_2) \to \langle U^\dag(t_2)
\rangle$ in the same manner as in Eq.~(\ref{c4})). Writing down the
sum of all products of traces of these three operators yields
\ba
\label{aa1}
&& \sum_k \frac{1}{2 \pi \rho_k^2 \Delta^2} \exp \{ - t^2 \Delta^2 /
(2 \hbar^2 \} \nonumber \\
&& \times \bigg( {\rm Tr}_k (A) \ {\rm Tr}_k \big( [\Pi A] (\Pi + \Pi^T)
\big) \nonumber \\
&& + {\rm Tr}_k (\Pi) \ {\rm Tr}_k \big( [\Pi A] (A + A^T) \big)
\nonumber \\
&& + {\rm Tr}_k \big( (A \Pi^T + A^T \Pi) ([\Pi A] + [\Pi  A]^T) \big)
\bigg) \ .
\ea
The factor ${\rm Tr}_k [\Pi A]$ occurs only if $U(t_2) \to \langle
U(t_2) \rangle$. That replacement is excluded because in the
third-order cumulant of $U(t)$ the factor $U(t_2)$ must be linked
either to $U(t_1)$ or to $U^\dag(t_1)$. As in Section~\ref{cor} the
matrix $[ \Pi A ]$ has elements $[ \Pi A ]_{m n} = \sum_l \Pi_{m l}
A_{l n}$ with an unrestricted summation over $l$. All remaining matrix
products and traces in expression~(\ref{a1}) are restricted to space
$k$. From the argument used to write down expression~(\ref{a1}) it is
not clear whether all terms in expression~(\ref{a1}) do indeed arise
nor which numerical factors they carry. That is irrelevant if it can
be shown that all terms in expression~(\ref{a1}) are negligible. While
${\rm Tr}_k (A)$ is of order $N_k$, the argument used in the first
paragraph of this Appendix shows that ${\rm Tr}_k \big( [\Pi A] (\Pi +
\Pi^T) \big)$ is of order unity. Combined with the factor $1 / (\rho_k
\Delta)^2$ the product of the two traces is negligible. With ${\rm
  Tr}_k (\Pi) = p_k$ of order unity and ${\rm Tr}_k \big( [\Pi A] (A +
A^T) \big)$ of order $N_k$, the product of the next two traces
combined with the factor $1 / (\rho_k \Delta)^2$ is negligible. The
remaining trace is at most of order $N_k$. Combined with the factor $1
/ (\rho_k \Delta)^2$ it is negligible. We conclude that contributions
of the third-order cumulant of $U$ to the correlation
function~(\ref{c1}) are negligible. As a check of our argument we have
calculated the contribution of the third-order cumulant of $U(t)$ to
the correlation function explicitly. That confirms
expression~(\ref{a1}).

We apply the argument of the last paragraph to the fourth-order
cumulant of $U$. Contraction of the matrix elements $O_{m \alpha}$
yields four Gaussian factors and four factors $1 / (\rho \Delta)$. The
expression is linked. Therefore, the energies in the exponential
factors of the four propagators $U, U^\dag$ all carry the same label
$\alpha$. The time dependence drops out. Summation over $\alpha$
yields a factor $(\rho \Delta)$ and confines the remaining four
summations to the same interval $k$. We are left with the sum over
products of traces of two operators $A$ and two operators $\Pi$, all
restricted to the same interval $k$. For brevity we write $\hat{A} =
(A + A^T)$, $\hat{\Pi} = (\Pi + \Pi^T)$. Writing down the sum of all
products of traces of these operators yields
\ba
\label{aa2}
&& \sum_k \frac{1}{(\sqrt{2 \pi} \rho_k \Delta)^3} \bigg( [{\rm Tr}_k
  (A)]^2 [{\rm Tr}_k (\Pi)]^2 \nonumber \\
&& + {\rm Tr}_k (\hat{A}^2) [{\rm Tr}_k (\Pi)]^2 + {\rm Tr}_k
      [(\hat{\Pi}^2] [{\rm Tr}_k (A)]^2 \nonumber \\
&& + {\rm Tr}_k (A) {\rm Tr}_k (\Pi) {\rm Tr}_k (A \hat{\Pi}) +
      {\rm Tr}_k [\hat{A}^2] {\rm Tr}_k [\hat{\Pi}^2] \nonumber\\
&& + [{\rm Tr}_k (\hat{A} \hat{\Pi})]^2 + {\rm Tr}_k (A) {\rm Tr}_k
      [(\hat{A}^2 \Pi] \nonumber \\
&& + {\rm Tr}_k (\Pi) {\rm Tr}_k [\hat{\Pi}^2 A] + {\rm Tr}_k [ \hat{A}
     \hat{\Pi} \hat{A} \hat{\Pi} + \ldots ] \bigg) \ .
\ea
The dots in the last trace indicate the sum over all permutations of
the four factors in the first term. We do not claim that all terms in
expression~(\ref{a2}) actually occur. But showing that all these terms
give negligible contributions implies that the terms that actually do
occur are negligible as well. Each trace is at most of order
$k$. Combined with the factor $1 / (\rho_k \Delta)^3$ that shows that
terms in expression~(\ref{a2}) which carry a single trace or two
traces, are negligible. In the terms carrying three traces, one trace
has either the form ${\rm Tr}_k \Pi = p_k$, or the form ${\rm Tr}_k
(\Pi)^2 < p_k$. Both traces are of order unity. Combined with the
factor $1 / (\rho_k \Delta)^3$, the terms carrying three traces are
negligible. In the term carrying four traces we also use ${\rm Tr}_k
\Pi = p_k$. The remaining two traces are of order $(\rho_k \Delta)$
each. Combined with the factor $1 / (\rho_k \Delta)^3$ they are
negligible. We conclude that contributions of the fourth-order
cumulant of $U$ to the correlation function~(\ref{c1}) are
negligible.

\end{document}